\documentstyle[pre,preprint,aps]{revtex}

\begin{document}

\title{Numerical study of a non-equilibrium interface model}

\author{Balakrishna Subramanian (sbala@physics.rutgers.edu)}
\address{Department of Physics, Rutgers University, Piscataway, NJ 08855, USA}

\author{G.T. Barkema (barkema@sns.ias.edu)}
\address{Institute for Advanced Study, Olden Lane, Princeton, NJ 08540, USA}

\author{J.L. Lebowitz\footnote{Also Department of
Physics, Rutgers University.}(lebowitz@math.rutgers.edu), E.R. Speer
(speer@math.rutgers.edu)} \address{Department of Mathematics, Rutgers
University, New Brunswick, NJ 08903, USA}

\maketitle

\begin{abstract}
We have carried out extensive computer simulations of one-dimensional
models related to the low noise (solid-on-solid) non-equilibrium
interface of a two dimensional  anchored Toom model with unbiased and
biased noise. For the unbiased case the computed fluctuations of
the interface in this limit provide new numerical evidence for the
logarithmic correction to the subnormal $L^{\frac{1}{2}}$ variance
which was predicted by the dynamic renormalization group calculations
on the modified Edwards-Wilkinson equation. In the biased case the
simulations are in close quantitative agreement with the predictions of
the Collective Variable Approximation (CVA), which gives the same
$L^{\frac{2}{3}}$ behavior of the variance as the KPZ equation.

\end{abstract}

\section{Introduction}
The nature of the interface separating two equilibrium phases, or more
generally any two distinct bulk states of matter, is a problem of continuing
interest. While there is in most cases some fuzziness in the
transition region, giving rise to an intrinsic structure, the width of this
is usually much smaller than (and therefore separable from) the fluctuations
in the location of the interface. These fluctuations are fairly well
understood both microscopically and macroscopically in (simple) equilibrium
systems \cite{DL}, but much less is known about them in the larger context of
nonequilibrium situations \cite{KS}. 

An important advance in the latter case was the introduction of equations
of the Edwards-Wilkinson \cite{EW} and Kardar-Parisi-Zhang \cite{KPZ} type
to describe fluctuations in dynamically moving interfaces.
Solutions of these equations appear to describe in a quantitative manner
the macroscopic fluctuations of a large class of such interfaces \cite{KS}.
It is probably fair to say, however, that there are few microscopic systems
(even simple ones) for which one can argue a priori, with mathematical rigor,
that their behavior will be described by these equations.

Several years ago two of us (JLL and ES), together with B. Derrida and
H. Spohn \cite{pap1,pap2}, introduced and studied the behavior of the
low-noise-limit sharp interface separating the $+$ and $-$ phases in a
simple model nonequilibrium system---the 2D Toom cellular automaton. Our
semi-infinite interface was situated in the third quadrant of the square
lattice and was anchored at the origin. Its precise location could be
specified by a spin configuration on the positive semi-infinite
one-dimensional integer lattice ${\bf Z}_+$. Fluctuations of the location
of the interface at a distance $L$ from the origin are then directly
transcribed into fluctuations of the magnetization $M_L= \sum_{i=1}^L
~\sigma_i$ in the stationary state of the one-dimensional model.

The dynamics of the semi-infinite 1D spin model are as follows: starting
with some initial configuration (containing a mixture of $+$ and $-$ spins)
we pick a site $i$ with rate 1 if $\sigma_i=-1$, and with rate $\lambda\leq
1$ if $\sigma_i=1$, and exchange $\sigma_i$ with the spin $\sigma_j$,
where $j$ is the first site to the right of $i$ such that
$\sigma_j=-\sigma_i$. Remarkably, the model inherits the unidirectional
nature of information flow in the Toom model and the stationary state for
the first $L$ spins can be obtained exactly (no finite size effects) by
restricting the dynamics to a system with L spins, with the provision that
if site $i$ is in the last block of spins, i.e., if
$\sigma_i=\sigma_{i+1}=\dots=\sigma_L$, then the spin $\sigma_i$ is flipped,
i.e., changed to $-\sigma_i$ (with rate 1 or $\lambda$). The model can
be classified into two cases, the unbiased case with $\lambda =1$, and the
biased case with $\lambda \neq 1$.

Quantities of interest in the model include the average of the magnetization
$\langle M_L\rangle$ and its variation
$ V_{L} \equiv \langle(M_L-\langle M_L\rangle)^2\rangle$ as functions of the
system size $L$. By symmetry, $\langle M_L\rangle=0$ in the unbiased case,
and a simple computation, using the well justified assumption that spins
far from the origin are statistically independent, shows that in general
$\langle M_L\rangle\simeq \mu L$ with
$\mu=\frac{1-\sqrt{\lambda}}{1+ \sqrt{\lambda}}$. The fluctuations have not
been calculated exactly. Simulations reported in \cite{pap1} gave
$V_{L} \sim L^\nu$ with $\nu=0.53$ in the unbiased case and $\nu=0.57$
for $\lambda=1/4$, but in fact approximate treatments discussed there
suggested that the true exponents are $\nu=1/2$ in the unbiased case, possibly
with logarithmic corrections, and $\nu=2/3$ in the biased case.

Two approximation methods were introduced in \cite{pap1}. The first, the
{\it collective variable approximation} (CVA), gives $V_L^{CV}\simeq
\sqrt{3/2}L^{1/2}$ in the unbiased case and $V_L^{CV}\simeq CL^{2/3}$ (with
$C$ a computable constant) in the biased case. The second approach was
based on a description of the interface by a non-linear stochastic
diffusion equation of the EW \cite{EW} and KPZ \cite{KPZ} type. Analysis of
this equation predicts how fluctuations of a (doubly) infinite uniform
interface will grow in time; in the semi-infinite problem, the fixed
boundary at the left of the system and the positive velocity of excitations
convert this growth to a corresponding $L$-dependence of $V_L$. This
approach predicts $V_L\sim L^{2/3}$ in the biased case (here $2/3$ is the
usual KPZ exponent). In the unbiased case the prediction depends on the
growth of excitations in a modified KPZ equation with a third order, but no
second order, nonlinearity; a subsequent renormalization group analysis of
this problem (\cite{BP}, see also \cite{DS}) leads to the prediction
$V_L\sim L^{1/2}\log^{1/4}(L/L_0)$.

The simulation results of \cite{pap1} did not include sufficiently large
systems to distinguish the logarithmic behavior in the unbiased case from
behavior $L^\nu$, with $\nu$ a power slightly higher than $1/2$, or to
verify the power $2/3$ in the biased case. Simulations of \cite{BP} and
\cite{DS} support the renormalization group conclusions for a uniform
system but, since the passage from the uniform to the semi-infinite system
is somewhat heuristic, provide only indirect guidance for the latter. In
the present note we describe new simulations on our original model for
$L\leq 2^{19}$ both for $\lambda=1$ and $\lambda=1/4$. We also did
simulations on modified models and models with different bias values. Our
results on the unbiased case give support to the predictions of \cite{BP}. Our
results for the biased case seem to suggest that the CVA is a very good
approximation.

For computational efficiency, we used a type of ``multi-spin coding'',
similar to the technique applied for simulating the repton model in
\cite{rep1,rep2}. The computationally intensive part of the code is written
in bit operations like AND ($\land$), exclusive OR ($\oplus$) and 
NOT ($\lnot$). This allows one to run 32 or 64 independent simulations
in parallel by applying bit operations to four-byte or eight-byte
integers of which each bit corresponds to an independent simulation.

We introduce a coding variable $\eta_{j}=(1+\sigma_{j})/2$. To implement
the algorithm, we first choose (as described below) a marker $m$, whose
nonzero bits select the simulations in which updating will take place, then
pick a random site $j$ and flip the selected spins at this site:
\begin{eqnarray}
\eta_j' & = & m \oplus \eta_j
\end{eqnarray}
Next we walk along the chain, starting at $ k=j+1$ and incrementing $k$ by one 
at each step. In each simulation we flip the spin $\sigma_{k}$ if it differs
from $\sigma_{j}$ and if so we set the corresponding bit in the marker to 
zero: 
\begin{eqnarray}
d & = & \eta_k \oplus \eta_j \\
\eta_k' & = & \eta_k \oplus ( m \land d) \\
m & = & m \land \lnot d
\end{eqnarray}
We continue walking until either our marker $m$ equals $\vec{0}$ or we
have reached the end of the chain. To insure that simulations running in
different bits do not become identical, we initialize $m$ as $m=r$,
where $r$ is a random bit-sequence with each bit taking the value 1 with
probability $1/2$, and then repeat with a different starting site $j$ and
with $m= \lnot r$. This generates a large degree of independence between
simulations corresponding to different bits.

To obtain the mask $m$ we use the random number generator {\it marsag}
\cite{marsaglia}, which has good randomness properties for all
the bits; for the site selection we use the random number generator
{\it ranmar} \cite{marsaglia}, which has a long sequence and good spectral
properties.

In the absence of a bias ($\lambda=1$), the thermalization time $t_{0}$
required was determined to be $t_{0}=L^2/8$ spin exchanges. The
autocorrelation time of the magnetization in the steady state is much
smaller, growing as $L^{3/2}$, as found also in \cite{pap1}, although
the thermalization time suggests that some correlations must grow as
$L^{2}$. The presence of a bias increases the required thermalization
time; for $\lambda=1/4$, $t_{0} \approx L^{2}/4$ and for
$\lambda=1/8,~t_{0}\approx L^{2}/2$.  For $\lambda=1/2$ we took again
$t_{0} \approx L^{2}/4$ as thermalization, although we could have taken
a bit less. One run typically starts with a random spin configuration,
which is evolved over time $(n+1)t_{0}$. The magnetization in each simulation
(corresponding to a bit) is measured every $L$ moves; for each interval
of length $t_{0}$ a separate histogram of the magnetization M,
including all bit simulations, is constructed. The first histogram is
discarded (thermalization) and from the remaining $n$ histograms we
obtain moments of the magnetization.  Reported statistical errors in
these moments are one standard deviation errors based on the assumption
that these $n$ histograms are statistically independent. Note that even
if simulations in different bits are correlated, we still obtain a
reliable error estimation.

\section{ The Unbiased Case}

In the absence of bias ($\lambda =1$) we now have data for sizes
$L \leq 2^{19}$, presented in Table I. Two analytic descriptions of the growth
of fluctuations have been proposed: in \cite{pap1} it was suggested that:
\begin{equation}
V_L \simeq C \cdot L^\nu
\label{power}
\end{equation}
with $\nu \simeq 0.5$; on the other hand in \cite{BP} (see also \cite{DS}), 
a formula based on the dynamical renormalization calculation was proposed:
\begin{equation}
V_L \simeq C \cdot L^{1/2} \cdot \log^{\beta}(L/L_{0}),
\label{log}
\end{equation}
with $\beta=1/4$ and $L_{0}=8$. In this section we will argue that this
data supports the conclusion of \cite{BP}, although we disagree 
on the identification of the constant $L_{0}$.

Clearly it is difficult to distinguish between the behavior (\ref{power}) and
(\ref{log}). For example, the quantity $V_L/(L^{1/2}\log L)$ is a
monotonic decreasing function of $L$ throughout the range of our
simulations and $V_L/(L^{1/2}\log^{1/8}L)$ is monotonic increasing for
$L\ge64$, suggesting asymptotic behavior (\ref{log}) for some intermediate
value of $\beta$, but in fact the same is true if $V_L$ is replaced by
$L^{0.53}$, the behavior found in \cite{pap1}.

As a qualitative criterion for comparison of fit, we ask for what minimal
value $2^{K}$ of the system size various proposed forms (all having two free
parameters) can provide a good fit over interval $[2^K,2^{19}]$, rejecting
a fit as bad if it fails the standard $\chi^2$ test at a 99\% confidence
level. The asymptotic form (\ref{power}), computed as a linear
fit $\log(V_L/L^{1/2})\sim a+b\log L$, yields $K=12$, while the forms
(\ref{log}), again as linear fits
$(V_L/L^{1/2})^{1/\beta}\sim a_\beta+b_\beta\log L$, yield $K=11$, $6$,
$8$, and $10$ for $\beta=1$, $1/2$, $1/4$, $1/6$, respectively. This
analysis thus provides evidence to prefer (\ref{log}), which describes the
data over a wider range of system sizes, although it does not conclusively
rule out (\ref{power}).

We now consider (\ref{log}) and ask whether our data can determine the
exponent $\beta$. The difficulty is that the three parameters in
(\ref{log}) provide a great deal of freedom, allowing for good fits for a
range of values of $ \beta$, if we vary the values of $L_{0}$ and $C$. The
test discussed above is indefinite, and in fact mildly favors a value of
$1/2$, as opposed to the renormalization group calculation of $1/4$. We now
turn to a different line of argument which does provide evidence for the
latter value, and begin with a review of the second approximate approach of
\cite{pap1}.

As mentioned in the introduction, the application of the renormalization
group depends on consideration of our dynamical model on the doubly
infinite line, viewed as a height interface growth model. Such interface
growth models may be described by non-linear stochastic diffusion equations
of the EW and KPZ type; for our model the symmetries in the problem lead to the
equation
\begin{equation}
\frac{dh}{dt} = \nu\frac{d^{2}h}{dx^{2}} + v_1\frac{dh}{dx} +
  v_3\left(\frac{dh}{dx}\right)^{3} + \eta(x,t) + \hbox{higher order terms},
\label{KPZ3}
\end{equation}
where $\eta$ is a stochastic noise term. 
The term in (\ref{KPZ3}) with coefficient $v_1$ is usually eliminated by a
Galilean transformation, but this is not possible in the semi-infinite
system and in fact this term has important consequences, discussed below.
If $v_3=0$ then (\ref{KPZ3}) is
the Edward-Wilkinson growth model, which leads to interface growth
$\langle h(t)^{2} \rangle \sim t^{\frac{1}{2}}$. The full equation
(\ref{KPZ3}) can be analyzed using dynamical renormalization techniques
\cite{BP}; in this analysis the higher order terms are irrelevant and the
cubic term is marginally irrelevant, giving
\begin{equation}
\langle h(t)^2\rangle \simeq At^{\frac{1}{2}} \log^{\frac{1}{4}}(t/t_0).
\label{time}
\end{equation}
The parameter $v_1$ is positive, so that fluctuations in $h(t)$ travel to
the right. We can compute the velocity $v$ of infinitesimal density
perturbations in the (Bernoulli) steady state, via
$v=\partial J/\partial \rho\big|_{\rho=1/2}$, where $J(\rho)$ is the
current at density $\rho$, to find $v=8$ \cite{BP}. To apply these
observations to the original model on the half open system it was suggested
in \cite{pap1} that excitations created at the closed end of the system
grow according to (\ref{time}) as they travel towards the open end at speed
$v$; this leads to (\ref{log}) with $\beta=1/4$, $C=A/v^{1/2}$, and
$L_0=vt_0$. (See also \cite{BP}, although there it is tacitly assumed that
$t_0=1$).

In order to understand the exact form of the logarithmic corrections and
check the validity of the above prediction, we studied a class of
modifications of the original model, with dynamics defined as follows: we
choose a site at rate $1$, then exchange the spin at that site with the
$k^{th}$ spin of opposite sign to the right. The original model is obtained
by taking $k=1$. These models all share the symmetries of the original,
so we expect that they will belong to the same universality class and hence
have similar behavior of fluctuations; thus the variance $V_{L}(k)$ in
these models should satisfy (\ref{log}) with the same value of $\beta$ but
with possibly different constants $L_{0}(k)$ and $C(k)$. We obtained data
for models with $k=2,3$ (see Table I). Again, we can fit (\ref{log}) to the
individual data sets for a range of values of $\beta$ if we choose
appropriate $L_0(k)$ and $C(k)$; to proceed further, we would like some a
priori argument determining how these values depend on $k$.

Now we propose a heuristic argument which describes how the constant
$C(k)$ might vary with $k$. On the one hand, a computation in the
steady state determines that the current in this model is proportional
to $k$ and hence that the velocity $v(k)$ of excitations satisfies
$v(k)=k \cdot v(1)=8k$.  On the other hand, the model for parameter $k$
can be thought of as a modification of the original model in which,
during one time step, $k$ exchanges rather than one spin exchange take
place.  Now these $k$ exchanges take place in correlated positions.
These correlations are certainly significant on the microscopic scale
(e.g., the distance between two successive moves would definitely be
affected by these correlations), but as seen from the computation of
the current and the velocity these correlations are apparently not
important on the hydrodynamic scale. Hence there should be some length
scale below which the correlations are significant.  We have two
parameters in (\ref{log}) which could be affected by these correlations,
$C(k)$ and $L_{0}(k)$.  $C(k)$ (along with $\beta$) determines the
behavior to leading order at large $L$, and $L_{0}(k)$ is
significant only at higher orders.  If we assume that the correlations
are not important at least to the order determining $C(k)$, then for a
computation of $C(k)$, the model for parameter $k$ would correspond to
the original model with the time rescaled by a factor of $k$. This
would imply that $A(k) \sim k^{1/2}A(1)$, and hence that
$C(k)=A(k)/v(k)^{1/2}=C(1)$.

Another way of saying this, which does not make any reference to the
time dependent problem, is to write $\langle M_{L+1}^2 \rangle =
\langle M_{L}^{2} \rangle + 2 \langle \sigma_{L+1}M_{L} \rangle + 1$.
Since $\langle M_{L+1}^2 \rangle - \langle M_{L}^{2} \rangle
\rightarrow 0$ as $L \rightarrow \infty$, we must have $\langle
\sigma_{L+1}M_{L} \rangle \rightarrow - \frac{1}{2} + o(L)$. Now if the
leading order term in $o(L)$ is also independent of $k$ (which seems not
unreasonable) then we would also have $C(k)$ independent of $k$ and vice
versa.

To study the $k$-dependence of $C(k)$, we plot in Figure 1
$(V_L(k)/L^{1/2})^{1/\beta}$ versus $\log{L}$ for $\beta=1/2$, $1/4$,
and $1/6$; according to equation (\ref{log}) this curve is approaching
a straight line with slope $C(k)^{1/\beta}$ and offset $C(k)^{1/\beta}
\log(L_0)$. Only for $\beta=1/4$ can a data collapse be obtained by
vertically shifting the curves (which corresponds to a change in
$L_0$), indicating that $k$-independence of $C(k)$ holds only for
$\beta=1/4$.  Results of least-squares fit for all data on the interval
$[2^{10},2^{18}]$, in the form $(V_L/L^{1/2})^{1/\beta}\sim
a_\beta+b_\beta\log L$, are given in Table 2; the approximate constancy
of $b_{1/4}(k)$ verifies the conclusion. Thus under the assumption made
above, this provides independent justification of the conclusion
$\beta=1/4$.

We also observe from Table 2 that $a_\beta(k)$ and hence $L_{0}(k)$ depends
strongly on $k$ for all $\beta$. We interpret this as indicating that the
correlations are important on the scale of $L_{0}$ and hence we cannot use
the above argument to determine the $k$ dependence of $L_{0}(k)$.

To summarize, we now have numerical evidence for logarithmic
corrections, and our data, when supplemented by a heuristic argument,
supports the prediction (\ref{log}) (with $\beta=1/4$) obtained with the
help of dynamical renormalization techniques \cite{BP}.

\section{ The Biased Case}

In the biased case we generated data for $L \leq 2^{18}$ for $\lambda=1/4$.
As compared to the data reported in \cite{pap1}, this provides much better
evidence for the asymptotic form $V_L \simeq BL^{2/3} $ predicted by both
the KPZ and CVA approximations. We note that for the biased model the
differential equation for the height evolution is the usual KPZ equation
\begin{equation}
\frac{dh}{dt} = \nu \frac{d^{2}h}{dx^{2}} + v_{1} \frac{dh}{dx} +
v_{2} \left(\frac{dh}{dx}\right)^{2} + \eta(x,t) + \hbox{higher order terms},
\end{equation}
where now the the non-linear quadratic term is relevant and it changes the
power of the growth of fluctuations in time from 1/2 to 2/3.

We now compare the simulation results more closely with those obtained
analytically for the asymptotic case of the CVA (the continuum limit):
$B^{CV} = 1.544 \lambda^{1/3} (1-\sqrt{\lambda})^{2/3}
(1+\sqrt{\lambda})^{-2}$ \cite{pap1}. Note that $B^{CV}$ vanishes at
$\lambda=0,1$ and has a maximum at $\lambda= 7-4\sqrt{3} \approx
0.0718$.
As shown in Figure 2, the values obtained from the CVA seem
to be strikingly close to the simulation values in the asymptotic
limit. The data suggests that the constants of proportionality $B$ for
the variance in the two cases are extremely close if not identical. To
check whether the CVA was close only for this particular value of
$\lambda$, we studied two more values of bias $\lambda =1/2,1/8$. As
shown in Figure 2 for both the cases, we observed that the CVA seemed
to give values which are extremely close to the actual values.

In order to have a better estimate of the accuracy of the CVA, we also
studied the whole distribution of magnetization. The continuum limit
CVA gives for the distribution of the magnetization values the fourth
power of an Airy function \cite{pap1}, with a somewhat ad hoc cutoff
suggested by the discrete CVA (the simulations and the CVA both give a
Gaussian distribution for $\lambda=1$). This function has an asymmetry
about the mean value. In Figure 3 we compare this CVA limiting
distribution with the distribution of magnetization determined from
simulations, with all distributions normalized to have a mean of 0 and
a standard deviation of 1. It can be seen that the actual distribution
shows the same characteristic features as the CVA result; moreover,
looking at the distributions for different L values suggests a slow
approach of the distribution towards the Airy function with increasing
size L. However, if we look at the leading measure of asymmetry in the
distribution--- the normalized third moment $\frac{\langle (M- \langle
M \rangle)^{3}\rangle}
      {\langle (M- \langle M \rangle)^{2} \rangle^{3/2}}$ --- then we
observe (see Figure 3 inset) that the actual values and that obtained
from the CVA are quite different.

We also studied the values of average magnetization obtained for finite
size systems. As noted in \cite{pap1} the exact asymptotic value and that
obtained from the CVA coincide and are given by $\langle M_L \rangle = \mu L$,
where $\mu=\frac{1-\sqrt{\lambda}}{1+ \sqrt{\lambda}}$. In Figure 4,
we plot the approach of the average magnetization to the asymptotic value.
Both the approach for the CVA and that for the actual data show very
similar behavior.

At the end of this analysis, we are left with the somewhat puzzling result,
that though the CVA does not reproduce, even qualitatively, the results for the unbiased case, it
seems to be an extremely accurate description for the biased case. 

{\bf Acknowledgments}

We thank T. Hwa, G. Sch\"utz and H. Spohn for useful discussion and
communications. GTB acknowledges financial support from the DOE under
grant DE-FG-90ER-40542, and from the Monell foundation. BS and JLL were
supported in part by NSF Grant 9213424. JLL would also like to thank
DIMACS and its supporting agencies the NSF under contract STC-91-19999 and the N.J. Commission on
Science and Technology.

\begin{figure} \caption{If $V_L \approx C(k) L^{1/2}
\log^\beta{(L/L_0)}$, then $(V_L/L^{1/2})^{1/\beta}$ as a function of
$\log{L}$ is a straight line with slope $C(k)^{1/\beta}$. In the
figure, $(V_L/L^{1/2})^{1/\beta}$ is plotted as a function of $\log{L}$
for $k=1$ (plusses), $k=2$ (circles), and $k=3$ (diamonds), and
$\beta=1/4$. The curves for $k=1$ and $k=2$ are shifted vertically, to
obtain a collapse. The insets show that for $\beta=1/2$ (left inset)
and $\beta=1/6$ (right inset) the data does not collapse.
\label{varbeta}} \end{figure}

\begin{figure} \caption{$V_L/L^{2/3}$ is plotted as a function of
$\log{L}$, for $\lambda=1/8$ (diamonds), $\lambda=1/4$ (circles), and
$\lambda=1/2$ (squares). In the first two cases the data shows
convergence to a constant, indicating that $V_L \sim L^{2/3}$ without
logarithmic corrections.  The lines are the CVA values for the
corresponding biases; the asymptotic CVA values are 0.3150, 0.2723 and
0.1855, respectively.
\label{bias}} \end{figure}

\begin{figure} \caption{Normalized distribution (with mean 0 and
standard deviation 1) of the magnetization $M_L$ for the biased case
($\lambda=1/4$) with system sizes $L=1024$ (dashed line) and $L=262144$
(dotted line) together with the normalized distribution obtained from
the asymptotic CVA (solid line). The inset shows the third moment of
the magnetization, as a function of $\log{L}$, for the CVA (solid line)
and for the simulation(circles).  \label{thirdmom}} \end{figure}

\begin{figure} \caption{Finite-size effects in the magnetization for
$\lambda=1/4$:  the deviation from $\langle M_L \rangle/L = 1/3$ versus
system size $L$ is given in a log-log plot. The solid line is the CVA,
the circles are simulation results.  \label{finite}} \end{figure}

\newpage
\newdimen\digitwidth\setbox0=\hbox{\rm0}\digitwidth=\wd0
\catcode`?=\active \def?{\kern\digitwidth}
\begin{table}
\renewcommand{\arraystretch}{0.7}

\begin{tabular}{rccc}
 $L$ & $V_L=V_L(1)$ & $V_L(2)$ & $V_L(3)$\\
\hline
2     & $1.335[1]$ & $2$ & $2$\\
4     & $1.918[1]$  & $2.75[1]$ & $3.637[1]$\\
8     & $2.733[1]$ & $3.818[1]$ & $4.626[2]$\\
16    & $3.890[1]$ & $5.271[1]$ & $6.451[2]$\\
32    & $5.576[2]$ & $7.338[3]$ & $8.868[6]$\\
64    & $8.02[2]$  & $10.286[4]$& $12.292[6]$\\
128   & $11.58[3]$ & $14.543[4]$& $17.20[1]$ \\
256   & $16.76[3]$ & $20.66[1]$ & $24.27[2]$ \\
512   & $24.22[4]$ & $29.44[2]$ & $34.37[2]$ \\
1024  & $35.01[3]$ & $41.96[6]$ & $48.70[3]$ \\
2048  & $50.54[9]$ & $59.94[8]$ & $69.20[3]$ \\
4096  & $73.1[1]$ & $85.6[2]$  & $98.41[4]$ \\
8192  & $105.3[2]$& $122.4[1]$ & $139.90[7]$ \\
16384 & $151.7[4]$ & $174.6[3]$ & $199.2[2]$ \\
32768 & $218.3[5]$ & $249.9[4]$ & $283.1[3]$ \\
65536 & $313[1]$ & $355.4[4] $ & $402.4[5] $ \\
131072& $449[2]$ & $507[3]  $ & $573[2] $ \\
262144& $644[4]$ & $720[2]  $ & $813[3] $ \\
524288& $924[3]$ &          & $1162[5] $\\ 
\end{tabular}
\vglue 1cm
\caption{ The variances obtained for the unbiased case for different
values of $k$. The quantity in brackets after each value is the
statistical error in the least significant digit.}

\newpage

\begin{tabular}{cccc}
 $k$ & $a_{1/2}$ & $a_{1/4}$ & $a_{1/6}$\\
\hline
 1 & $0.694\pm0.018$ & $0.086\pm0.050$ & \llap{$-$}$0.969\pm0.102$\\
 2 & $1.376\pm0.023$ & $1.682\pm0.084$ &  $1.551\pm0.228$\\
 3 & $2.052\pm0.018$ & $4.104\pm0.084$ &  $7.924\pm0.300$\\
\hline
\hline
 $k$ & $b_{1/2}$ & $b_{1/4}$ & $b_{1/6}$\\
\hline
 1 & $0.073\pm0.002$ & $0.194\pm0.006$ & $0.386\pm0.013$\\
 2 & $0.050\pm0.003$ & $0.184\pm0.009$ & $0.507\pm0.025$\\
 3 & $0.038\pm0.002$ & $0.179\pm0.010$ & $0.640\pm0.036$\\
\end{tabular}
\vglue 1cm
\caption{Values of $a_\beta$ and $b_\beta$ obtained from least-squares fits on
the data for the unbiased model.}

\newpage
\begin{tabular}{rccccc}
$L$ & $V_L$ & $ \langle M \rangle_L$ &
 $\langle (M-<M>)^3 \rangle_L$ & $V_L$
 & $V_L$\\
&$\lambda=\frac{1}{4}$ & $\lambda=\frac{1}{4}$& $\lambda=\frac{1}{4}$ &
$ \lambda=\frac{1}{2}$ & $\lambda=\frac{1}{8} $ \\
\hline
1024 &  36.82[6] & 344.57[2] &  $-$0.171[6] & 35.7[1] & 37.0[2]\\
2048 & 55.3[1] & 686.98[2] & $-$0.185[2] & 52.4[2] & 57.2[3] \\
4096 & 83.4[2] & 1371.06[3] & $-$0.204[4] & 76.4[4] & 88.5[6] \\
8192 & 128.8[4] & 2738.32[5] & $-$0.225[3] & 111.9[5] & 136.6[8] \\
16384 & 193[2] & 5471.4[1] & $-$0.240[3] & 165.5[6] & 215[2] \\
32768 & 302.7[6] & 10936.1[3] & $-$0.256[2] & 247[1] & 344[3] \\
65536 & 473[1] & 21862.62[6] & $-$0.266[4] & 370[5] & 533[5] \\
131072 & 735[4] &  43713.2[2] & $-$0.281[6] & 566[5] & 849[10] \\
262144 & 1165[7] &  84105.9[2] & $-$0.266[5] &  & \\
\end{tabular}
\vglue 1cm
\caption{The variances for three different values
 of bias,  and the first and third moments for $ \lambda
=\frac{1}{4}$. Note that we have not obtained data for systems of size
smaller than $1024$.  The statistical error in the least significant
digit is given by the quantity
 in brackets.}

\end{table}
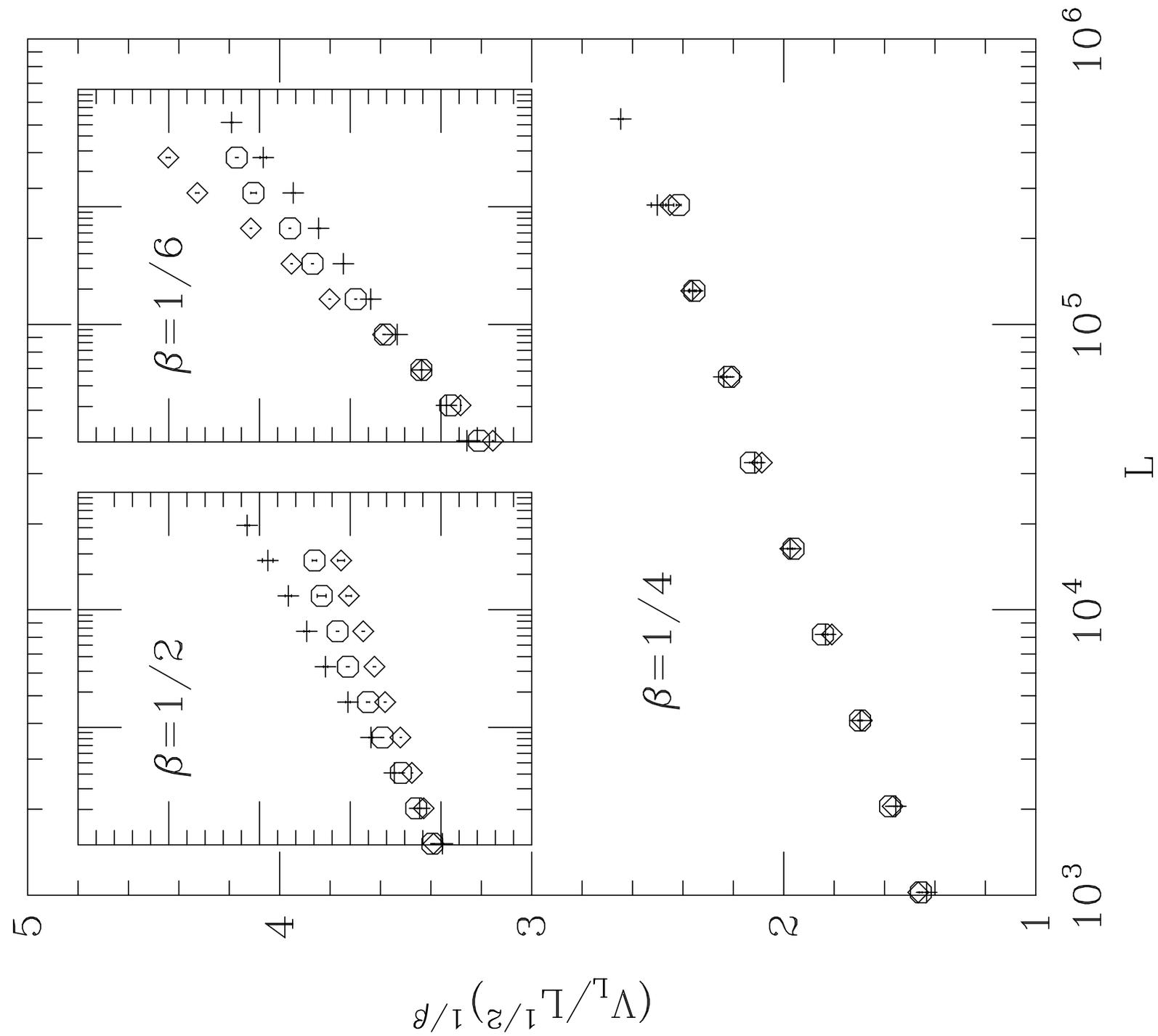
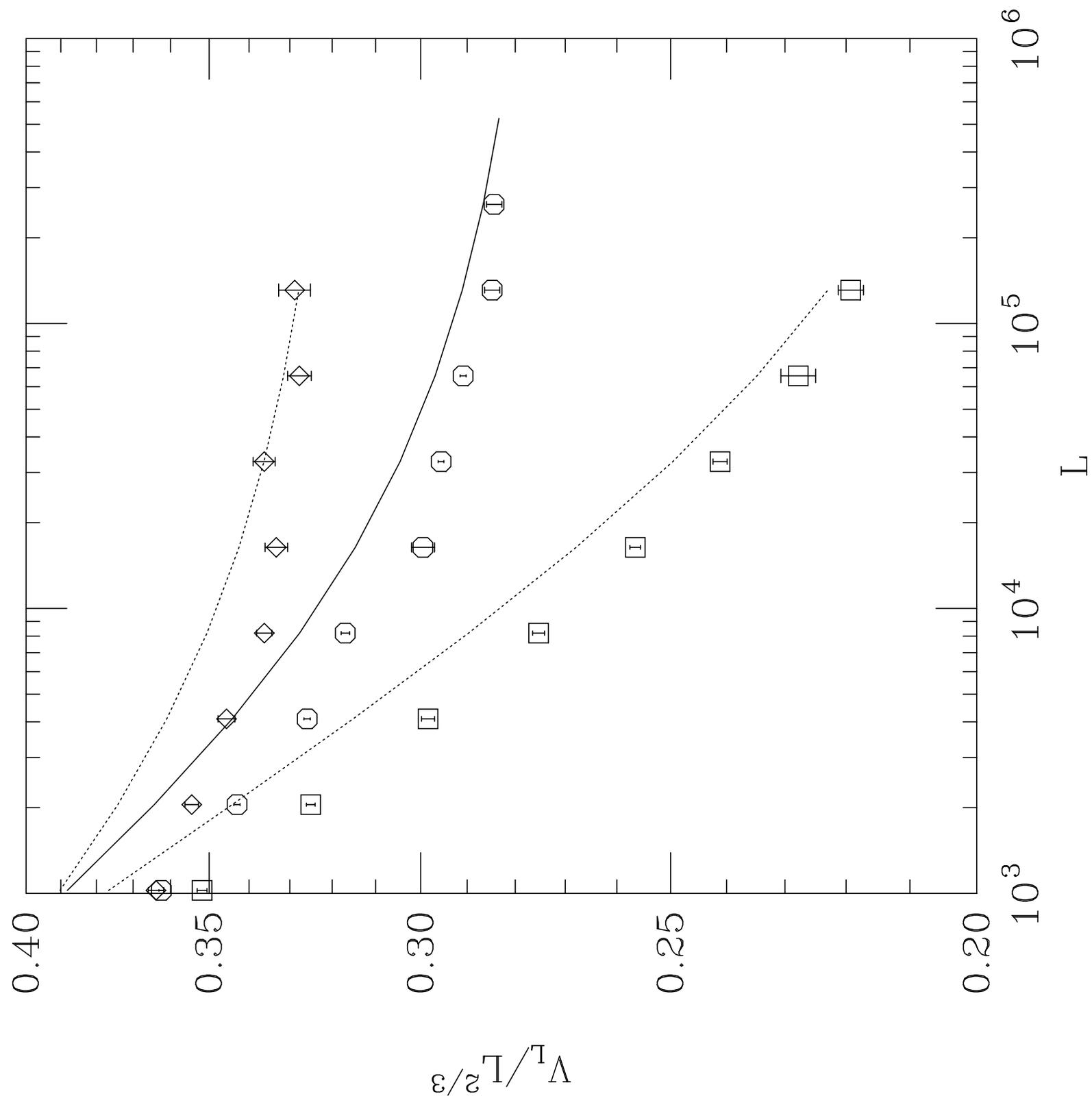
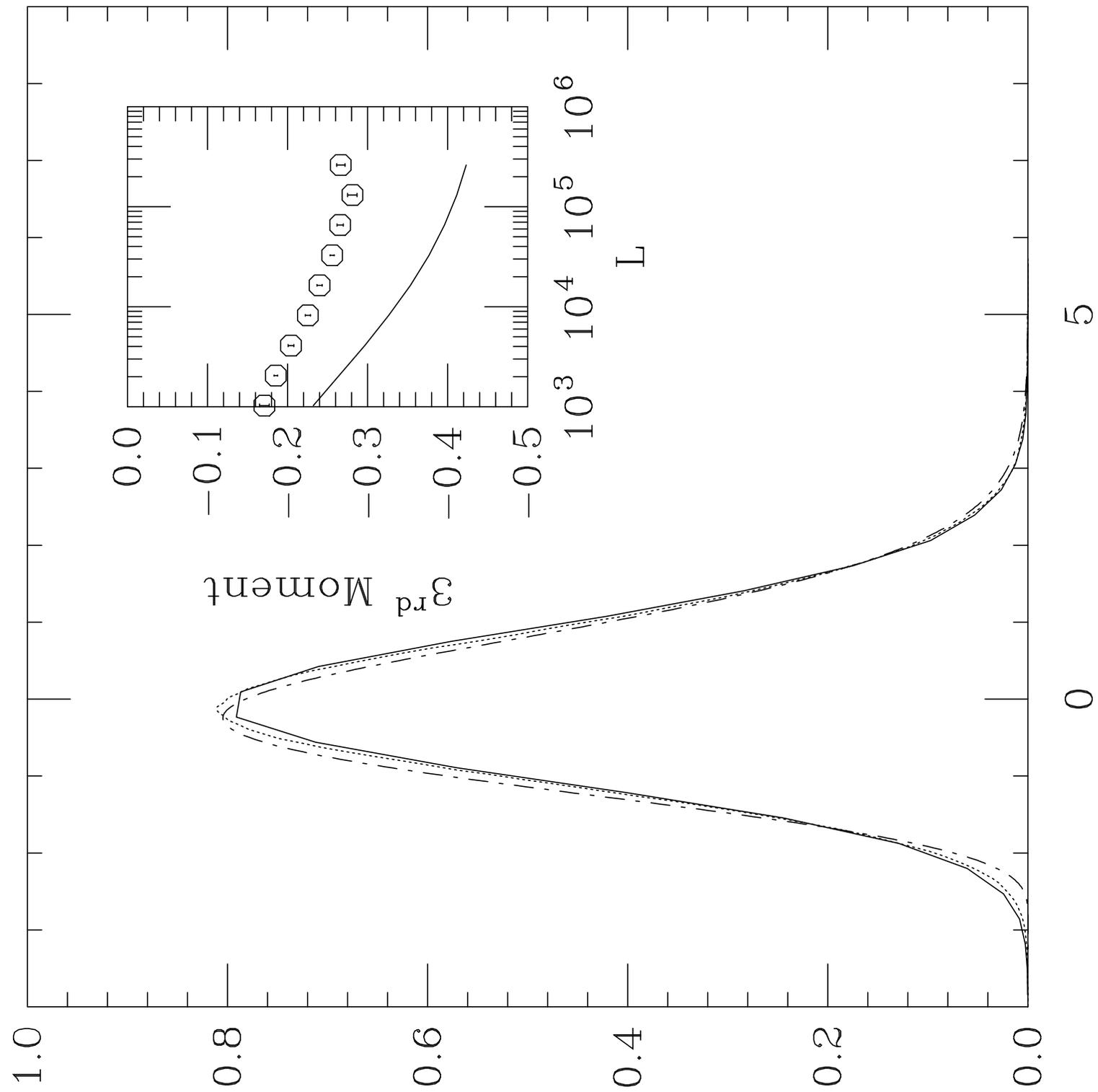
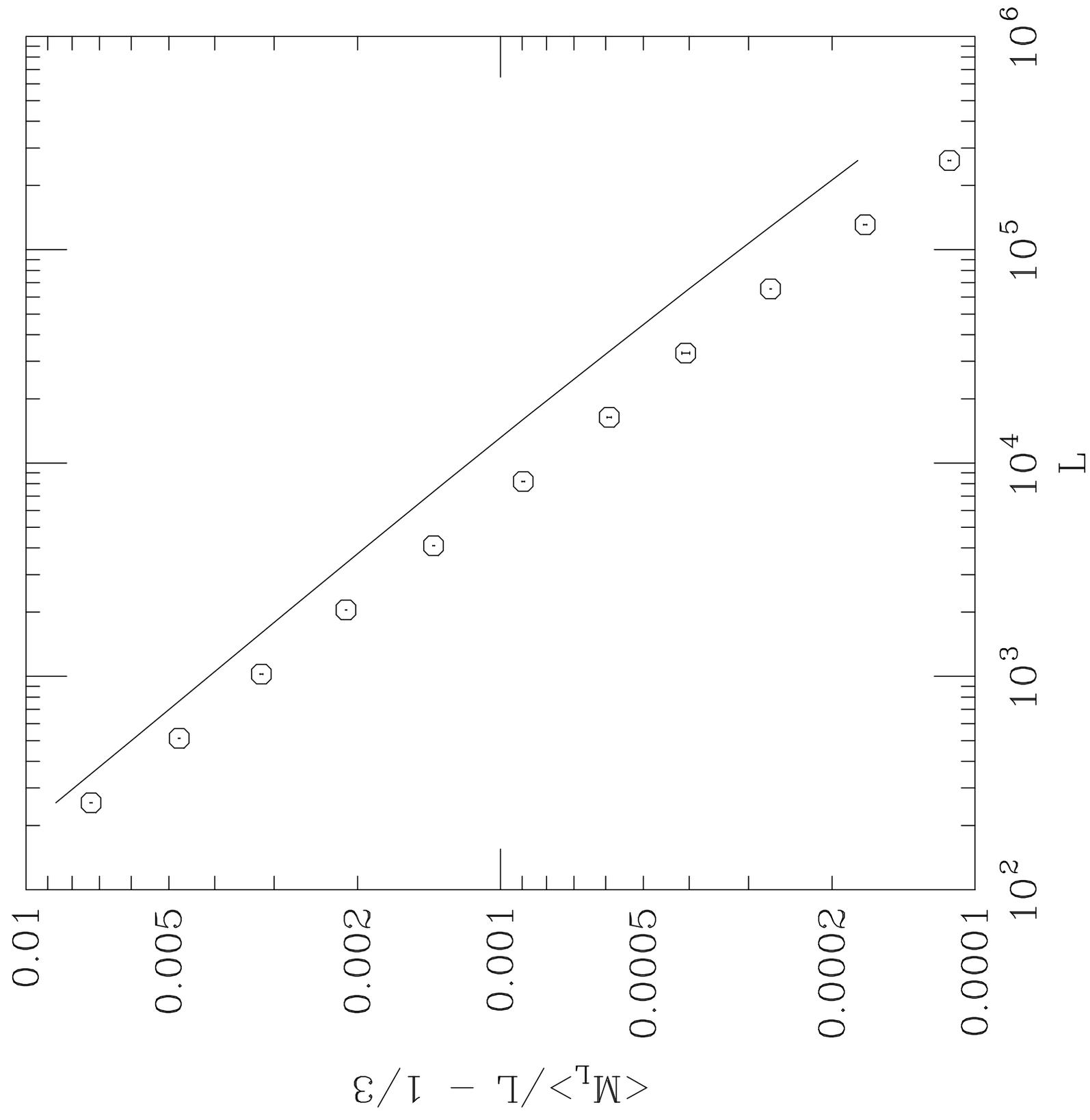
\end{document}